\documentclass[12pt]{iopart}
\usepackage{graphicx,psfrag}
\usepackage[latin1]{inputenc}
\newcommand{\ds}{\displaystyle}
\newcommand{\mb}{\mathbf}
\newcommand{\notes}[1]{}

\newcommand{\ep}{\epsilon}
\newcommand{\beq}{\begin{equation}}
\newcommand{\eeq}{\end{equation}}
\newcommand{\beqas}{\begin{eqnarray*}}
\newcommand{\eeqas}{\end{eqnarray*}}
\newcommand{\beqa}{\begin{eqnarray}}
\newcommand{\eeqa}{\end{eqnarray}}
\newcommand{\p}{\partial}

\begin{document}

\title[Spin-wave interference for memory and computation]{Spin-wave interference patterns created by spin-torque nano-oscillators for memory and computation}
\author{F. Maci\`a$^*$}
\address{Department of Physics, New York University, 4 Washington Place, New York, NY 10003}
\address{Courant Institute of Mathematical Sciences, New York University, 251 Mercer Street, New York, NY 10012}
\ead{fmb2@nyu.edu}
\author{A.D. Kent}
\address{Department of Physics, New York University, 4 Washington Place, New York, NY 10003}
\author{F.C. Hoppensteadt}
\address{Courant Institute of Mathematical Sciences, New York University, 251 Mercer Street, New York, NY 10012}
\date{\today}

\begin{abstract}
Magnetization dynamics in nanomagnets has attracted broad interest since it was predicted that a dc-current flowing through a thin magnetic layer can create spin-wave excitations. These excitations are due to spin-momentum transfer, a transfer of spin angular momentum between conduction electrons and the background magnetization, that enables new types of information processing. Here we show how arrays of spin-torque nano-oscillators (STNO) can create propagating spin-wave interference patterns of use for memory and computation. Memristic transponders distributed on the thin film respond to threshold tunnel magnetoresistance (TMR) values thereby detecting the spin-waves and creating new excitation patterns. We show how groups of transponders create resonant (reverberating) spin-wave interference patterns that may be used for polychronous wave computation of arithmetic and boolean functions and information storage.
\end{abstract}

\maketitle
\section{Introduction}
Polychronous Wavefront Computation (PWC) as a paradigm for a post digital era was introduced by Izhikevich and Hoppensteadt in 2009 \cite{izhifhijbc} following the early work of Izhikevich \cite{izhipwc}, where the importance of spatial propagation and axonal delays in the brain were investigated. The key ingredients for PWC are a medium that can support interference patterns of propagating activity packets, and transponders that can sense the size of incident activity and respond to super-threshold inputs by generating a propagating wave. Mathematical models that support such features are two dimensional wave equations and Schrödinger's equation in bounded domains with absorbing boundary conditions. Transponders are devices located at specific locations that can receive and transmit activity when appropriately stimulated. PWC is inspired by studies of the brain by Izhikevich \cite{izhipwc}, where it was shown that delays caused by axonal conduction times introduce new possibilities for network behavior, including memory and computation. These are referred to as being \textit{polychronous} aspects of brain function.

Here we describe how to construct devices that use polychronization arrays of spin-torque nano-oscillators (STNOs) in two-dimension ferromagnetic thin films. The propagation times of peak wave interference correspond to conduction-time delays. The analogy of excitation of a neuron is excitation of a transponder that may fire when a certain threshold of inputs from other transponders is exceeded. In this case, the transponder generates a new wave, perhaps after some time delay. The work presented here is not an attempt to model a brain, but to accomplish some brain-like behaviors on much shorter time scales and on smaller spatial scales. Our devices support the propagation of activity patterns on spatial scales of nanometers at gigahertz frequencies, while brain activity occurs on millimeter scales at tens-of-hertz frequencies.

\subsection{Wavefront Intersections}

Associated with a wave is a cone in space-time whose cross-section expands in time proportional to the wave speed. If the region is isotropic, the cones are right cones. If the region is irregular, the cones are irregular, as well. Energy packets may not be uniformly distributed on the cone, as we demonstrate later for spin waves. The surface of a cone will be referred to as being a wave front along which energy packets may propagate. Interference patterns created by energy packets on cones reveal the possibilities for PWC and guides programming. In some applications, energy is distributed uniformly on the cone (e.g., the excitation produced by a single STNO diffuses isotropically). In others, energy packets, while always moving on the surface of the cone, may not be uniformly distributed on it (e.g., certain excitations of some STNO arrangements produce interference patterns and consequently irregular energy distributions by means the space-time cones).

\begin{figure}[htp!!]
\centering
\includegraphics[width=100mm]{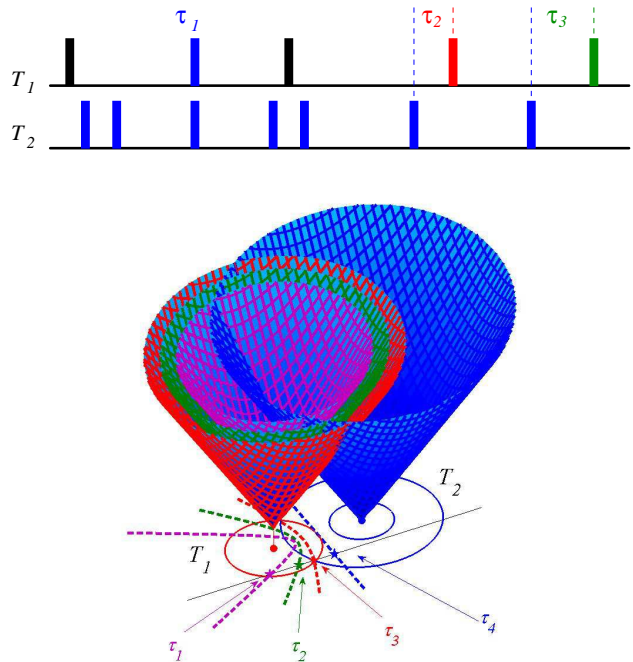}
\caption{Transponders $T_1$ and $T_2$ fire with a range of time delays. The waves will intersect along the curves indicated in dashed lines projected in to the $t=0$ plane. Transponders labeled with stars detect and eventually fire whenever there is an inter-pulse interval of length $\tau_i$, respectively; the array processes signals from transponders $T_1$ and $T_2$ by converting temporal delays into spatial firing patterns.}
\label{fig1supl}
\end{figure}

The notion of signal conduction delay has a clear representation in interference patterns created by waves. Two transponders firing at different times may excite propagating energy packets that may be visualized as lying on cones in space-time. A cone will begin at the location and time of initial excitation, and expand outward with increasing time. Two cones may intersect in a conic section, usually in a parabolic curve. This curve describes the evolution of the intersection of the two waves. The parabolic-shaped curve on which wavefronts intersect will be referred to here as being an intersection parabola as shown in Fig.\ \ref{fig1supl}. (If one cone is contained within the other, there will be no intersection.) A transponder lying on an intersection parabola may be excited by the wavefronts converging on it. Transponders may be placed at differing distances from a common target transponder. Given that each wavefront has the same finite propagation times, delays are encoded in Euclidean distances. We also allow transponders to be excited externally.  We can therefore introduce additional time delays into our system simply by exciting two transponders at different times.

Complicated polychronous activity is coded in PWC by the relative positions of transponders, and in the timing of external excitations. We illustrate this idea in Fig.\ \ref{fig1supl} by describing possible curves of intersection of pulses emanating from two separate transponders $T_1$ and $T_2$ generated at separate times, say $\tau$ units apart.

Several aspects of programming for PWC are comparable to sound location and response mechanisms in animal brain. For a given task, we show how to arrange transponders to accomplish it. This mode of programming is comparable to learning in the human brain. For example, multiplication tables are \emph{learned} by shaping our neural network to provide a look-up table mechanisms for the result of multiplying two numbers. One aspect of the work here describes how to create look-up table arrays using PWC. Other aspects regard storage of computations, detection of inter-wavefront intervals as a means for recovering stored information, and possible physical mechanisms for implementation.

It is shown in \cite{izhifhijbc,narendra} how to place transponders to accomplish computations such as addition, subtraction, multiplication and division.

\section{STNO design}
The interaction of a spin-polarized current with a magnetic film results in magneto-resistance \cite{fert,grunberg} and a spin-transfer-torque effect \cite{slonczewski,berger,slonczewski2,Tsoi,Kiselev}. The latter consists of a torque exerted by the spin-polarized electrons on the background magnetization. Promising applications based on the spin transfer effect are spin-torque nanoscillators (STNOs) \cite{Tsoi,Kiselev,Rippard} where spin-transfer-torque is used to drive a GHz oscillation of the magnetization direction of the magnetic film. STNO consisting of a point contact to a thin film ferromagnet (FM), were first proposed theoretically in 1996 \cite{slonczewski,berger}. DC current densities ($j > j_{\mbox{c}}$) generate a high-frequency dynamic response (1 to 100 GHz) in the FM layer and result in the emission of spin-waves \cite{slonczewski2}.
%Studies of STNO to date have relied primarily on electronic transport characteristics, i.e., spin-waves are studied indirectly through the effect that magnetization changes have on contact $I - V$ characteristics \cite{rippard}. Further studies have shown that these oscillations may be phase-locked to an external radio-frequency (rf) source \cite{rippard2}, via a process known as injection locking, providing means to phase and frequency lock an array of STNO. Moreover, proximal STNO experimentally exhibit phase locking, and the coupling mechanism is thought to be spin-wave radiation in a common FM layer \cite{kaka,mancoff}.

\begin{figure}[ht]
\centering
\includegraphics[width=100mm]{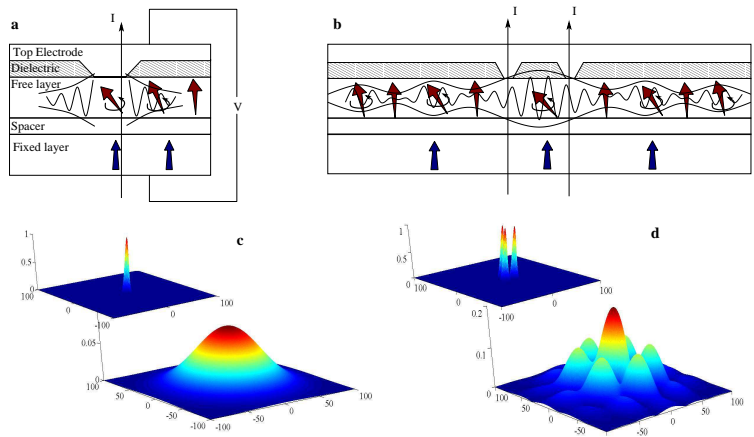}
\caption{Schematic of an STNO. (a) A single STNO: the charge flows through a point contact to a thin ferromagnetic layer. The magnetic moments in the contact area precess, exciting spin-waves that diffuse in the film. (b) A double STNO arrangement is shown. The double point contact creates a spin-wave interference pattern. (c) The spatial dependence of the amplitude of the excitation $|m|^2=m_x^2+m_y^2$ is plotted as well as one of the in-plane components, $m_x$. (d) Three point-contacts in a triangular arrangement are excited with a current pulse producing a magnetic excitation; the magnetic excitations diffuse and interfere throughout the two-dimensional thin film. Intersecting waves create six wave-packets that propagate in different directions. Simulations are performed in a 500 by 500 nm film with point-contacts of 20 nm diameter, and separation between contacts of 60 nm.}
\label{fig1}
\end{figure}

The basic concept of STNO devices is illustrated in Fig.\ \ref{fig1}(a). A nanoscale electrical contact is attached to a multilayered ferromagnetic structure \cite{Tsoi,Kiselev,Rippard}. The multilayer consists of a fixed and a free magnetic layer separated by a non-magnetic metal. The fixed layer determines the polarization of the current that excites magnetic moments in the free layer. A typical fixed layer would be a 10 nm thick permalloy film (with in-plane magnetization) or a multilayered structure (with an out-of-plane magnetization). The free layer may be a much thinner film (1-3 nm) \cite{SloncPatent,Kent}. The contact sizes range from tens to hundreds of nanometers. Phase locking between oscillators has been observed for distances up to 500 nanometers \cite{kaka,mancoff,grollier}, and the spin waves are predicted to propagate distances of tens of microns.

The magnetic moments of the free layer initially precess around the direction of the applied field and eventually align with the field as the result of damping. The response to a (polarized) current in a free magnetic thin layer is phenomenologically described with an additional current-dependent positive damping term in the Landau-Lifshitz-Gilbert equation. The result is a microwave excitation of magnetic moments of the thin film layer. Such excitations have a precession frequency due to the local magnetic field that depends non-linearly on the amplitude of the excitation. If we consider a unit vector to represent the direction of magnetization, $\tilde{m}$, then in the geometry considered in Fig.\ \ref{fig1} the out-of-plane magnetic moment $m_z$ does not oscillate and it represents the envelope of the modulated in-plane components, $m_x$ and $m_y$, ($m_z^2=1-(m_x^2+m_x^2)=1-m^2$, we call $m=(m_x,m_y)$ the in-plane component for convenience). Weak excitations at a single contact decay radially (see, Fig.\ \ref{fig1}(c)).

Micromagnetic modeling on shaped magnetic thin-film waveguides \cite{Choi} and on double coupled STNOs \cite{Choi2} has been done, and interference patterns have been found \cite{Lee2,Bance}. Arrangements of more than one STNO can create interference and enhance activity in some preferred directions in the film's plane. The richness of the magnetic-moment dynamics during the dissipation of initial pulses is indeed enormous and it motivated our methodology for generating interference patterns throughout the magnetic layer. A spatial pattern created by three STNOs arranged in the vertices of a triangle is depicted in Fig.\ \ref{fig1}(d).

Multiple current pulses excite multiple spin-waves that interfere. These create patterns in the envelope of the magnetic moments (i.e., $m_z$ for our geometry), and the constructive interference of wave-packets from several STNOs creates peaks in particular radial directions. By engineering both the distance between contacts and their sizes, it is possible to control spin-wave patterns. Notice that excitations must be phase locked within the STNO arrangements; an external background signal might be used to lock in phase or in frequency the STNOs \cite{kaka,mancoff,Rippard2005,Pufall,bonin,zhou,Rezende,slavinieee2005,slavin}. In addition, the frequency modes in the background signal might contribute energy for information storage \cite{FHactivity} (see our discussion in section\ \ref{sync}).

\section{Mathematical Theory}

Mathematical modeling is based on the Landau-Lifshitz-Gilbert equation in thin ferromagnetic films with an additional positive damping term corresponding to the transferred torque from the polarized current to the film. Let $\mb{M}$ denote the magnetization vector in the thin film, then if an external magnetic field is applied, the spin-magnetic moments will initially precess around the direction of the applied field, and eventually align with the field as the result of damping. An applied polarized current will cause a positive damping, and if the applied current exceeds a certain threshold ($j>j_{\mbox{c}}$), then a spin-wave excitation will be created \cite{slonczewski,berger,slonczewski2,Tsoi,Kiselev}.
This is described by the equation
\beq
\frac{\p \mb{M}}{\p t}=-|\gamma| \mu_0 \mb{M} \times \mb{H_{eff}}-\alpha\mb{M} \times (\mb{M} \times \mb{H_{eff}})+ \beta(\mb{x}) (\mb{M}\times \mb{M} \times \mb{m_p}),
\label{lle}
\eeq
where the precession (first term) and damping (second term) are controlled by the effective field $\mb{H_{eff}}$, being the sum of the applied field, the demagnetizing field, and the exchange field,
\beq
\mb{H_{eff}}=H_0\mb{z}-M_z\mb{z}+\frac{D}{|\gamma|\mu_0 M_s \hbar}\nabla^2\mb{M}.
\eeq

The spin-torque (third term) is controlled by the spin polarization direction of the applied current, $\mb{m}_p$. No variations in $z$ are considered (the Laplacian is taken in two dimensions), as the free layer is thin with respect to the exchange length. The function $\beta(\mb{x})$ is a Heaviside
function defining the contact sizes and locations. Furthermore, the function $\beta$ depends on the current intensity, the layer thickness and the spin polarization.

We define dimensionless parameters
\beq
\mb{m}=\frac{\mb{M}}{M_s}=(m_x,m_y,m_z),\ \ t=\frac{\omega_M}{2\pi}\tau, \ \ \mb{x'}=\frac{\mb{x}}{l_{\mbox{ex}}},
\eeq
where $\omega_M= 2 \pi \gamma \mu_0M_s$ is the Larmor frequency for an applied field $M_s$ and $l_{\mbox{ex}}=\sqrt{D/\gamma\mu_0M_s\hbar}$ is the exchange length.

Since the magnetization vector $\mb{m}$ lies on the unit sphere and the motions of interest are small deviations from equilibrium, we consider the components of $\mb{m}$ as
$$
m=m_x+\imath m_y, m_z=\sqrt{1-|m|^2},
$$

With these variables the Landau-Lifshitz equation takes the form of a nonlinear Schr\"odinger equation

\beq
\fl
\begin{array}{rl}
\ds \imath\frac{\p m}{\p \tau}=& -m\nabla^2m_z+ (m_z+\imath\alpha')\nabla^2m-(m_z-h)m
-\imath\alpha'(h-m_z)m_zm+\imath\alpha'\left(|\nabla m_z|^2+ |\nabla m|^2\right)m\\\\
&+\imath j\Phi \biggl[ \bigl(m_zm\bigr) m_{F_z} + \bigl(m_xm_y-(m_z^2+m_y^2)\bigr)m_{F_x}
+\bigl(m_xm_y-(m_z^2+m_x^2)\bigr)m_{F_y}\biggr] \\\\
\ds \frac{\p m_z}{\p \tau}=& -\mbox{Im}(m^*\nabla^2m)+ \alpha'(h-m_z)|m|^2
+\alpha'\left[\nabla^2m_z+(|\nabla m|^2+|\nabla m_z|^2)m_z\right]\\\\
&-j\Phi \biggl[|m|^2m_{F_z}-(m_x m_z) m_{F_x}+(m_y m_z)
m_{F_y}\biggr],
\end{array}
\label{eq:schrodinger}
\eeq
where $\nabla^2$ denotes the 2-dimensional Laplacian and $\alpha'$ is the (normalized) damping. The dimensionless parameters have roughly the following values: For a permalloy film the saturation magnetization is about 640 kA/m. Thus, the Larmor frequency, $\omega_M$, is about $2\pi\cdot23$ GHz, the exchange length, $l_{\mbox{ex}}$, is about 6 nanometers, and the dimensionless damping, $\alpha'$ is of order $10^{-2}$.

Applying a small-parameter perturbation analysis ($m=a\tilde{m}$ for a $a \ll 1$) and linearizing Eq.\ \ref{eq:schrodinger}, one gets
\beq
\imath\frac{\p \tilde{m}}{\p t}=
(1+\imath\alpha')\nabla^2\tilde{m}-(h-1)\tilde{m}-\imath\alpha(h-1)\tilde{m}.
\eeq
With the linearized equation, the evolution of excitations at different point contacts can be added and patterns studied without integrating the full equations of motion (eq.\ \ref{eq:schrodinger}). We present, however, in Sec.\ \ref{simulation} computations using the full LLG equations (Eqs.\ \ref{eq:schrodinger}) that corroborate this analysis.

\section{Transponder design}
A transponder is a device that can detect spin-wave activity in the propagating medium and respond by initiating new spin-waves. It is basically an integrate-and-fire circuit \cite{izhibook,frank}, where a storage device (e.g., a capacitor or an inductor) accumulates charge that is discharged rapidly when a certain threshold level is achieved. Such devices are widely used in electronics and physics ranging from the macroscopic neon bulb \cite{neon} to nano-sized memristors \cite{Chua,Chua2,memfound,yang,Borghetti,Borghetti2}. A basic design for a transponder that we have implemented in computer simulations is described here.

\begin{figure}[ht]
\centering
\includegraphics[width=100mm]{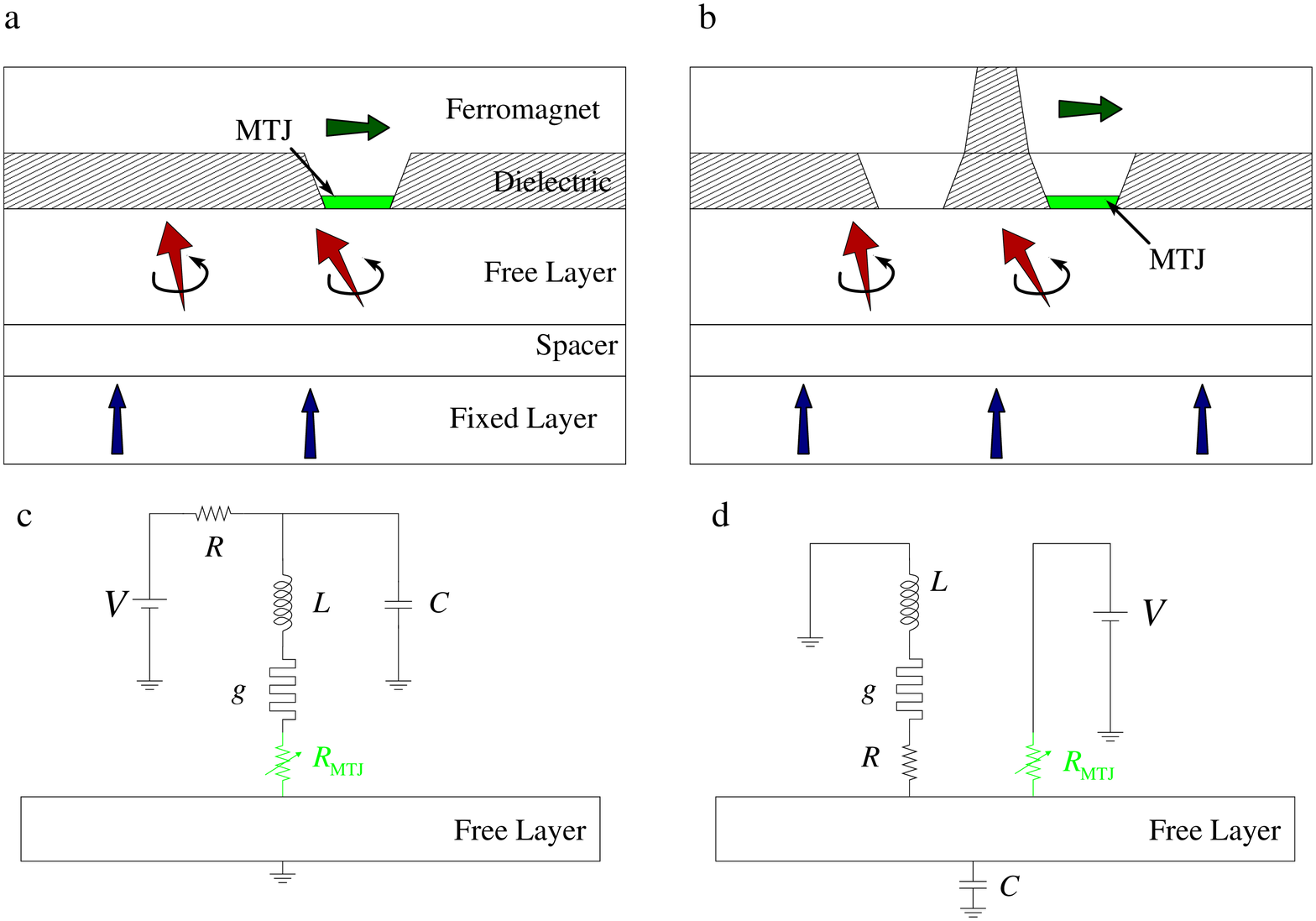} %\end{center}
\caption{A schematic representation of a tunnel junction on top of the contact of the STNO in (a) and next to the excitable point contact in (b). An additional ferromagnet layer is required to detect tunnel magnetoresistance (TMR) in both cases. $V$ is the applied voltage representing a dc-bias, $R$ is a resistor, $V_c$ is the voltage across the capacitor, $L$ is an inductor and $R_{\mbox{MTJ}}$ is the resistance of the the tunnel junction. In (c), $(V-V_c)/R$ provides a current source, and it splits into the current into the capacitor ($C(dV/dt)$) and the current $I$ through the point contact (in the resting state the current through the capacitor is 0). In (d), the current source is given by  $(V-V_c)/R_{MTJ}$. In both cases $R_{MTJ}$ controls the voltage across the memristor and whether it stays in the low or high-resistance branch of its $I-V$ characteristics curve.}
\label{fig4supl}
\end{figure}

STNOs are used to excite the magnetic moments in thin films and the same (or other) contacts can also be used as detectors. The respective alignment of the fixed and free layers determines the resistance of the STNO. Thus, small currents (that do not excite magnetization dynamics) through the point contacts would serve to read the state of the free layer. However, the giant magnetoresistance effect is weak when the orientation of one layer tilts only a few degrees, particularly when the layer magnetizations are initially collinear. If we include an additional dielectric layer on top of the STNO we obtain a magnetic tunnel junction (MTJ), which is either in series or parallel with the contact. A ferromagnetic top electrode (with the magnetization orthogonal to the free layer) provides the detector with a variable resistance depending on the orientation between the magnetic moments of the free layer and the magnetization of the ferromagnet on top (see, Fig.\ \ref{fig4supl}(a) and (b)).

A memristic element together with a capacitor, an inductor and an additional resistance are part of the detector as well. Thin films of TiO$_2$ may be deposited on top of the MTJ and may even serve as the dielectric in the MTJ. Memristors (memory resistors) \cite{Chua,Chua2} are any passive two-terminal circuit elements that maintain a functional (nonlinear) relationship between charges and magnetic flux. There are no generic memristors. Instead, each device implements a time-varying function of net charge history. Our circuit will includes additional elements to enable current bias bistability in the STNO and the integrate and fire functionality (see Figs. \ref{fig4supl}(c) and (d)).

Point contacts have a very low resistance ($R\sim 1 \ \Omega$) and the required current densities to excite spin-waves lead to tiny heat dissipation. On the other hand, MTJ have resistance of hundreds of ohms and consequently the same current densities would produce considerable greater dissipation. Figure\ \ref{fig4supl}(b) shows a parallel configuration of STNO plus MTJ. Additionally, Figs.\ \ref{fig4supl}(c) and d depict detecting schemes corresponding to both the in-series and parallel configurations of the detectors.

\begin{figure}[ht]
\centering
\includegraphics[width=100mm]{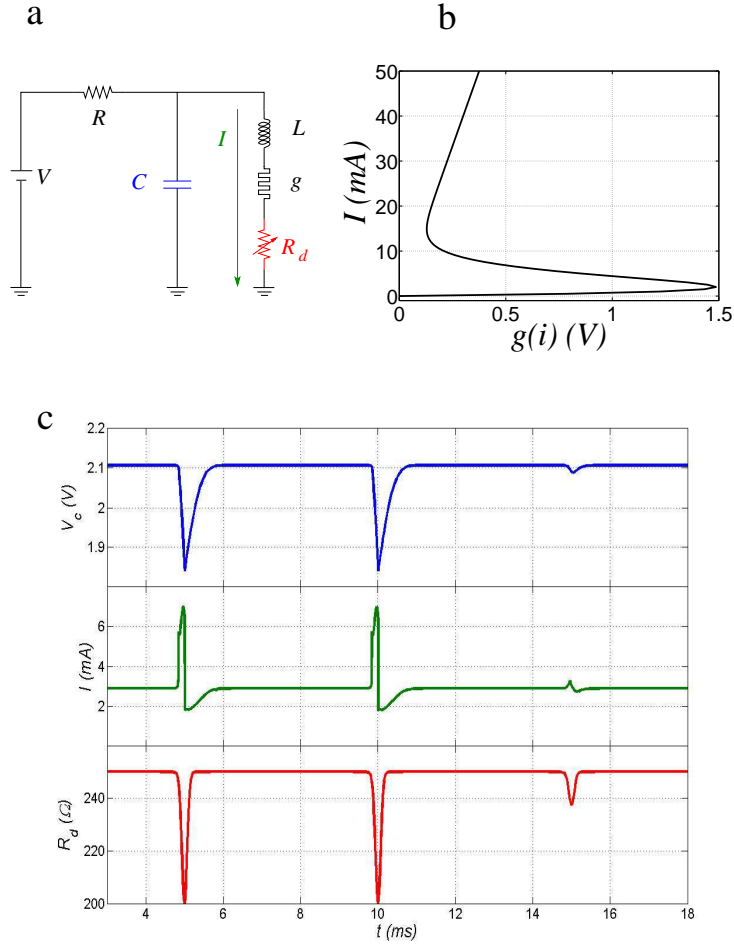}
\caption{An implementation of a super-threshold memristic circuit in (a). $V$ is the applied voltage representing a DC-bias, $R$ is a resistor, $V_c$ is the voltage across the capacitor, $L$ is an inductor and $R_d$ is the resistance of the point contact plus the tunnel junction. $(V-V_c)/R$ provides a current source, and it is split into the current into the capacitor ($C(dV/dt)$) and the current $I$ through the point contact (in the resting state the current through the capacitor is 0). The voltage drop across the memristic element is $V_g = L(d I/dt) + g(I)+IR_d$. In (b) a simulation of the circuit in (a). This figure present the time dependence of both time the current through the point contact and the voltage in the capacitor. The resistance $R_d$ is the input signal, and it varies in time (bottom panel) and produces discharges of the capacitor through the memristic element (it varied three times but only the first two exceeded the threshold and produced current discharges).}
\label{fig3}
\end{figure}

The process of detection is as follows: 1. A small voltage applied to the detector point contact induces a small current (insufficient to produce magnetic excitations in the free layer, $j<j_c$), that is used for detecting the TMR through the tunnel junction. 2. The capacitor $C$ is charged and the memristic device is in its high resistance branch. 3. When a spin wave perturbation reaches the point contact detector, its resistance changes ($R_d(\theta)$). 4. The change in resistance produces a voltage change in $g$ with time constant ($RC$). 5. When the threshold in $g$ is reached, there is a jump from the high-resistance branch to the low-resistance one and the capacitor, $C$, is discharged through the point contact, thereby producing a pulse excitation. 6. Once the capacitor is discharged the resistance at $g$ resets to the high-resistance branch and the charging process restarts. 7. During the charging, the value of $R_d$ is no longer important. No detection is possible until the capacitor is charged.

The equations describing such detection schemes are
\noindent
\beq
\begin{array}{rcl}
\ds C\frac{\partial V}{\partial t}&=&(S-V_c)/R_1-I
\\\\
\ds L\frac{\partial I}{\partial t}&=&V_c-g(I)-R_2-I,
\end{array}
\label{cireq}
\eeq
where $R_1$ and $R_2$ will either be the resistance of the MTJ, $R_{MTJ}$, or the fixed resistance, $R$.
The function $V_c=g(I)$, is the $I-V$-characteristic given by the function \cite{neon} (see Fig.\ \ref{fig3}(b))
\beq
g(I)=0.0075I+2I\exp(-I/2)
\eeq
(The I-V characteristics of Pt$|$TiO$_2|$Pt \cite{memfound,yang,Borghetti,Borghetti2} or phase change memory cells would also work).

A simulation of the super-threshold memristic circuit is included in Fig.\ \ref{fig3}(c). $V$ is the constant applied voltage representing a dc-bias. The voltage at the capacitor, $V_c$, together with the current $I$ across the STO are plotted as a function of time. The resistance of the MTJ is forced to vary at $t=5,10$ and 15 ns. The forced drops of $R_{MTJ}$ cause a shift in the memristor $I-V$ curve and eventually the resistance jumps from the high-resistance branch to the lower-resistance one, and at that moment the capacitor is discharged through the STNO and the initial state in the memristor (high-resistance branch) is rapidly recovered and the capacitor recharged again. Notice that small variations of the TMR do not produce a change of state of the memristor and consequently the capacitor is not discharged through the STNO (that is the case of the variation at $t=15$ ms).

If two transponders (i.e., two arrangements of STNOs) spike together, the excitation wave-packets will intersect at different times and positions. These locations and times can be determined using the geometry of spin-wave intersections. For example, the intersection of two wave fronts will occur on a parabola-like curve, and any transponder placed on this curve may be used to detect the event. Notice that delays in the excitation of the two spin-wave fronts result in shifts of the spin-waves intersection curves so the delays may be detected by differently located transponders.
Super-threshold detectors would be set to detect only excitations caused by more than one transponder, thus, protecting the system against other perturbations $-$such as thermal noise. Differential detection comparing the TMR signal in differently located transponders can also be used to determine the spin-wave patterns.

\section{Analysis and Simulation of the Patterning}
\label{simulation}

We first demonstrate interference patterns between spin waves emanating from two sources (two point contacts) \cite{Tsoi,Kiselev,Rippard} using numerical simulations of Eq.\ (\ref{eq:schrodinger}). Note that the waves add as they pass each other, and the intersections trace out parabolic curves. Thus, the coincidence of two spin-waves should be detectable by a transponder in the ferromagnetic film placed along an intersection parabola. Numerical simulations indicate that at the convergence of two waves the magnetization vector is deflected approximately twice that when a single wave passes.
\begin{figure}[htp!!]
\centering
\includegraphics[width=100 mm]{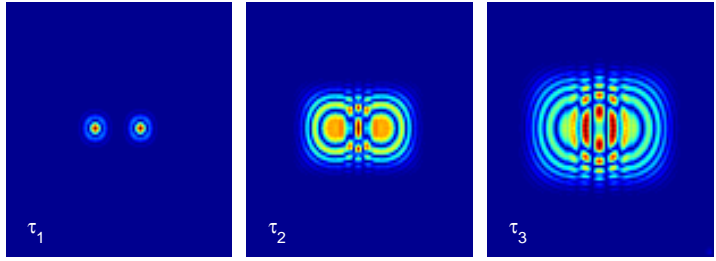}
\caption{Spatial dependence of the in-plane component of the magnetization, $m_x$, at 3 different times, $\tau_1 < \tau_2 < \tau_3$. A pair of point contacts are initially excited by polarized currents. The excitations diffuse throughout the two-dimensional film. Interference patterns are also created in $m_y$. Note that the intersecting waves pass through each other, and their amplitudes add. Simulations are performed in a 440 by 440 nm film, contacts of 20 nm in diameter are separated by 125 nm. The current-polarized pulses were switched on for $50$ ps and the depicted times correspond to $\tau_1=15$ ps, $\tau_3=175$ ps and $\tau_3=265$ ps after the polarized pulse was switched off.}
\label{fig2supl}
\end{figure}

Figure\ \ref{fig2supl} shows spatial dependence of one of the in-plane components of the magnetization ($|m_x|$ for convenience), at 3 different times, $\tau_1<\tau_2<\tau_3$, of the waves emanating from two STNOs after being excited with current pulses.

\begin{figure}[htp!!]
\centering
\includegraphics[width=100 mm]{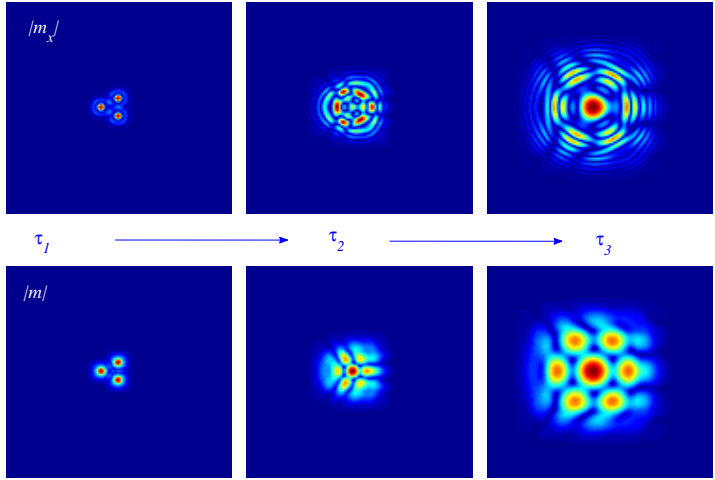}
\caption{Three point contacts create magnetic excitations that diffuse and interfere throughout the two-dimensional thin film. Intersecting waves create six wave packets, and they propagate outwards in specific directions. The upper panels show the evolution of the amplitude of one of the in-plane components, $|m_x|$; the lower, show the evolution of the perturbation amplitude, $|m|$. Three different times are plotted, $\tau_1 < \tau_2 < \tau_3$. Simulations are performed in a 440 by 440 nm film, contacts of 20 nm diameter, and separation between contacts of 65 nm. The current-polarized pulses were on for $50$ ps and the depicted times correspond to $\tau_1=15$ ps, $\tau_2=175$ ps and $\tau_3=265$ ps after the polarized pulse was switched off.}
\label{fig3supl}
\end{figure}

Note that the out-of-plane magnetic moment $m_z$ represents the envelope of the modulated, in-plane components, $m_z^2=1-(m_x^2+m_y^2)=1-|m|^2$, and single contact excitations decay radially. In addition to the spatial patterning shown in Fig.\ \ref{fig2supl} that diffuses, the frequency, $\propto|\gamma|H_0$, cause the in-plane components, $m_x$ and $m_y$ to oscillate. Thus, encoding information in the envelope (i.e., magnitude of the oscillation, $|m|$) creates spatial patterns that do not oscillate in time and only diffuse/propagate. Here we considered our transponders to be a group of point contacts.

Figure\ \ref{fig3supl} shows the interference patterns for both the amplitude of one of the in-plane components, $m_x$, and the amplitude of the perturbation, $|m|$, created by 3 identical current-pulses applied to a triangular array of STNOs. In this particular case the excitations interfere in such a way that the spin-wave are enhanced at the vertices of an expanding hexagon. The complexity of the in-plane components, $m_x$ and $m_y$ is enormous as they oscillate at a fast rate compared with the diffusion times. The envelope, $|m|$, captures the essence of the perturbation, and indeed reflects those areas where excitations are; zeros in the the envelope means no spin-wave activity but zeros in either $m_x$ or $m_y$ carry no information on the intensity of the spin-wave excitation.

\section{Reverberating Structures}
The next step is to design a configuration of transponders in which data can be stored. There are several configurations that can maintain stable reverberating activity \cite{izhifhijbc}, and hence serve as memory storage units.

\begin{figure}[ht]
\centering
\includegraphics[width=100mm]{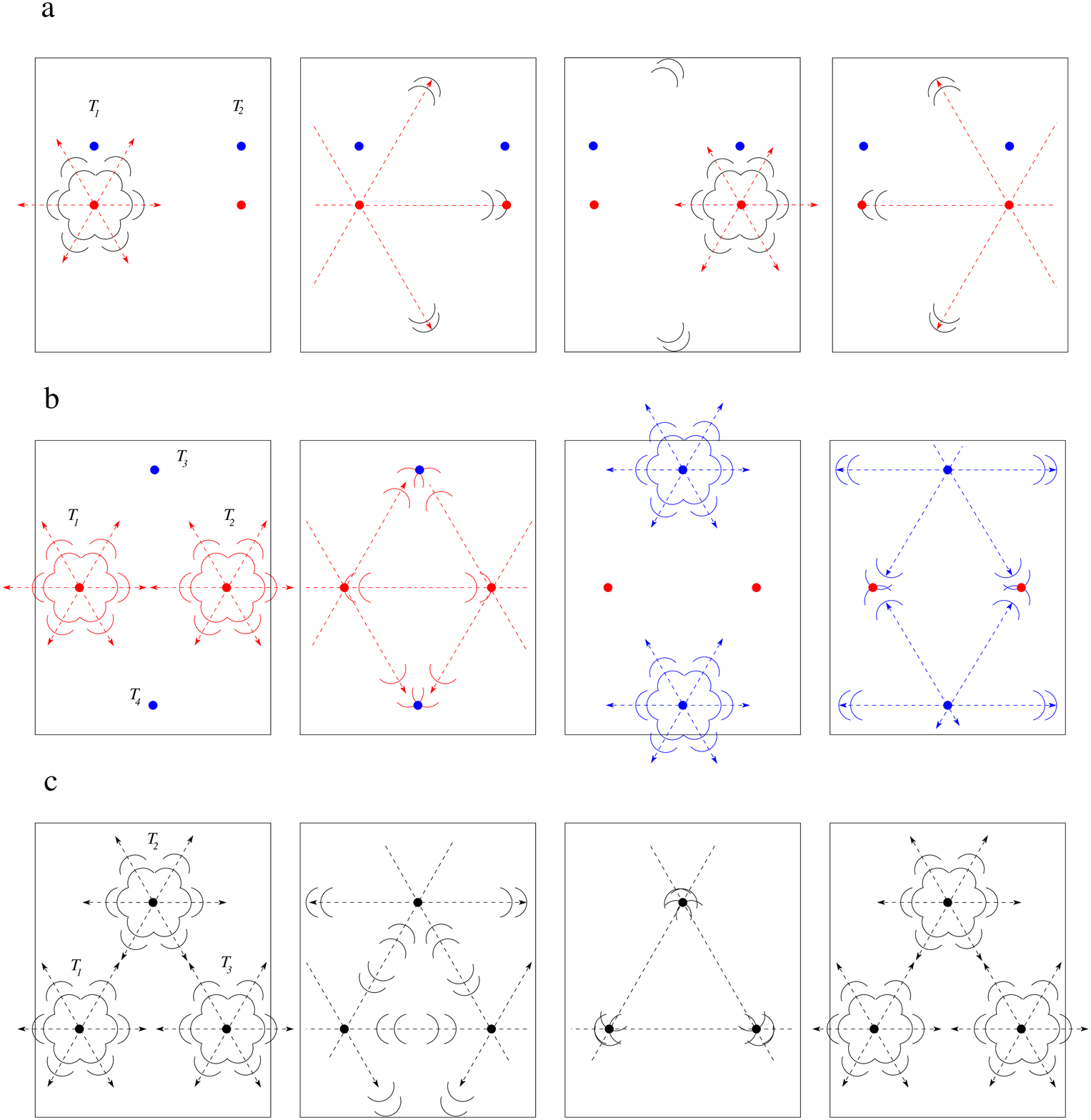}
\caption{In (a), differential detection configuration. In (b) and (c) configurations requiring converging  and intersecting wave-packets.}
\label{fig4}
\end{figure}

Combinations of transponders (number and location) result in controllable interference patterns, but these patterns may be unrecognizable from single-point readings. Spatial information may be obtained with detectors in order to distinguish whether the detected perturbation was created in, say position $a$, or position $b$. We note that should single detectors be used, several point contacts will have to fire simultaneously in different locations through the film and create constructive patterns in the detector that surpass its threshold; which would be set to respond only to higher amplitudes larger than that of one excitation. In Fig.\ \ref{fig4}(a) transponders $T_1$ and $T_2$ are made of two sets of point contacts, one red and one blue. The red one in $T_1$ spikes as in Fig.\ \ref{fig3supl} propagating the perturbation along three directions. The transponder $T_2$ receives a wave-packet in the red detector and nothing in the blue one; only this configuration allows $T_2$ to induce a new excitation that will propagate again through the media and be detected by $T_1$.

Another simple configuration that reverberates is when transponders $T_1$ and $T_2$ in Fig.\ 4b are excited at the same time, then wave-front packets propagate from $T_1$ and $T_2$ intersect at $T_3$ and $T_4$, therefore, the wave-front packets pass through $T_3$ and $T_4$ at the same time exciting a new pair of waves, which return to $T_1$ and $T_2$ at the same time, etc. The reverberation loop is closed and the activity is periodic with a period depending on $(D+D')/2$, where $D$ is the distance between $T_1$ and $T_2$, and $D'$ is the distance between $T_3$ and $T_4$.

Interestingly, one can implement similar reverberating memory using three transponders (no differential detection is required here), as in Figure\ 4c.

\subsection{Simulation of a reverberating structures}
\begin{figure}[htp!!]
\centering
\includegraphics[width=\columnwidth]{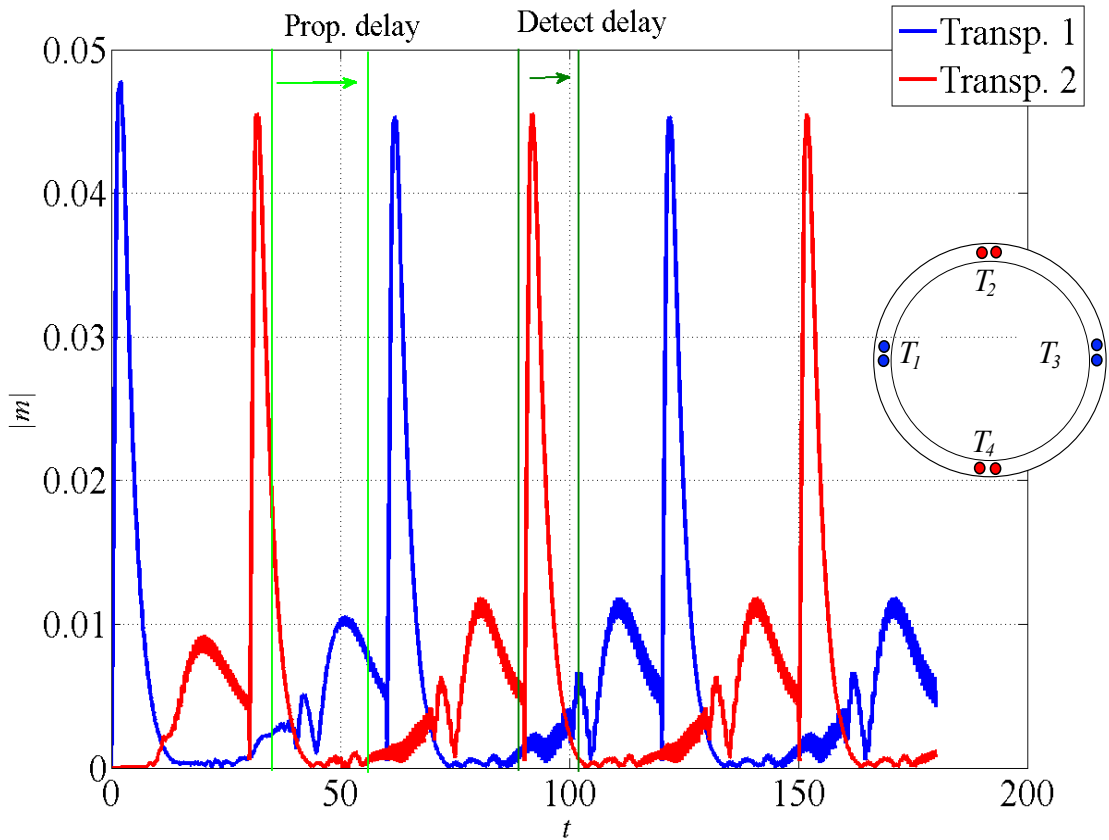}
\caption{Time evolution of the envelope, $|m|$, at transponders $T_1$ and $T_2$. At zero time, $T_1$ and $T_3$ are forced to spike together (positive current during 1 time unit) creating a considerable perturbation to the in-plane component of the magnetic moment. The perturbation propagates towards the neighbors. The in-plane magnetic moment in transponders $T_2$ and $T_4$ increases reaching a maxima after 20 time units. Once the incoming wave packet is detected $T_2$ and $T_4$ spike together (positive current for 1 time unit) and the pattern keeps repeating. }
\label{fg:circular}
\end{figure}

%\subsection{Transponder reverberation}
Following is a simulation of a four-transponder reverberating system; four double transponders, are arranged in a one dimensional ring as shown in the inset of Fig.\ref{fg:circular}.

Transponders $T_1$ and $T_3$ initially spike together. The spin-wave perturbations diffuse and propagate through transponders $T_2$ and $T_4$ where their interference pattern will be detected by those transponders ($T_2$ and $T_4$). The variation of the magnetic moments are detected, and after a time delay both transponders ($T_2$ and $T_4$) spike together and create a diffusing spin-wave perturbation that propagates towards $T_1$ and $T_3$. The cycle is closed once the transponders $T_1$ and $T_3$ detect the perturbation and spike again after a time delay. This reverberation may continue for a long time unless a powerful enough noise source appears and interferes in the reverberative configuration. Figure \ref{fg:circular} depicts the time evolution of the absolute value of the in-plane magnetization, $|m|$, at transponders $T_1$ and $T_2$ for a simulation of eq.\ \ref{eq:schrodinger}.

\section{Discussion and Conclusions}

It was shown in Ref.\ \cite{izhifhijbc,narendra} how reverberating structures may be used to perform Boolean and arithmetic computations. Any medium supporting steady progressing waves, vortices, or solitons may be used to implement PWC.

%The methodologies described above may be implemented in other physical systems in a number of ways.

To add more flexibility to the model, we allow for multiple layers of the two dimensional structures just described, which may share transponders. Layers could communicate via shared transponders and have, at the same time, non-shared transponders. In other words, if a transponder on one sheet is excited, it may or may not be allowed to produce propagating waves from the equivalent position on a nearby layer. Moreover, we may place additional transponders in this second layer wherever we choose. One can imagine multiple layers laid on top of each other or arranged in more complex topologies. Such designs remain to be investigated.

An application would be programming transponder arrangements having graded reverberating frequencies to create a look-up table, similar to ones in the brain. Consider the analogy of a piano keyboard; an array of several devices as in fig\ \ref{fig4}(c) having a gradient of natural frequencies (i.e., different separations between nodes in reverberators result in different resonant frequencies), are simultaneously exposed to a double-spike signal with period $\tau$. The device with resonant interspike interval $\tau$ will respond coherently, while the others will not. Namely, the double spike will not match with the resonant frequency of others and will create complex reverberating activity from which the coherent one can be distinguished. Programming by transponder placement is similar to learning in the brain where connections may be strengthen by training.

Reverberation may be controlled, if desired, by modulating any one of the nodes or a common oscillating background signal. It is also possible that a reverberator could be interrupted by interference patterns created accidentally from other systems, and there may be interference from itself if significant noise or boundary reflections are present. For the most part, the computations presented here do not exhibit undesirable interference patterns since the calculations are all synchronous and require a threshold for recognition, i.e., either two waves converging on a transponder or two transponders detecting an interference pattern. Damping of signals, and the introduction of absorbing boundary conditions may limit the impact of wavefronts on distant configurations, but these aspects as well as the impact of random noise on such reverberators require further analysis.

The PWC methodologies have been shown to enable arithmetic and boolean operations and look-up table functionality. Only the latter has been discussed here. Other computational aspects of the implementations described in this paper and the precision of PWC, which, for example, is limited by the breadth of wavefronts, will be the subject of future work. Methodologies shown in this paper focused on spin-waves created by STNOs. Other nano-systems may also support wavefronts in service of PWC, such are arrays of phase-locked magnetic vortices \cite{Fertvortices}, arrays of Josephson junctions \cite{Likharev}, and rapid single flux quantum devices \cite{imn2,likharev}. Systems that support solitons of activity may also provide useful implementations.

%Systems other than those described here may also support wavefronts in service of PWC such as arrays of phase-locked loops, arrays of Josephson junctions, and rapid single flux quantum devices \cite{imn2,likharev}. Other systems that may support solitons of activity may also provide useful implementations.

\section*{Acknowledgements}
Supported in part by FENA/FCRP Grant No. 6160-G-FD211 and by ARO-MURI, Grant No. W911NF-08-1-0317. FM acknowledges support from a Beatriu de Pinós grant from the Catalan Government.

\appendix

\section{A Note on Synchronization and Background Oscillations}
\label{sync}

It may be desirable in most applications to provide the system with an external background oscillation to ensure the interference occurs between point contacts having exactly the same driven frequency and phase. Several modes of oscillation (i.e., several distinct frequencies) can be considered \cite{FHactivity}. The background excitations can be created either with additional persistent ac-currents applied equally to all contacts or directly applying microwave radiation to the whole system of contacts \cite{Rippard2005,Pufall,bonin,zhou,Rezende,slavinieee2005,slavin}.

The equation describing the precession amplitude, $z(t)$, of an STNO is reduced to the normal form of a Hopf Bifurcation \cite{slavin,frank}
\beq
\frac{dz}{dt}= (\alpha-|z|^2)z + \imath \omega z,
\eeq
where $\alpha$ is now a coefficient involving both positive and negative damping.

With some mathematical treatment it can be shown there exist intervals around the oscillating frequency, $\omega$, where the oscillators phase-lock either to an external frequency or to another oscillator.

\subsection{Phase-Locking of STNOs to External Sources}
Consider an STNO forced by an external signal having several frequencies, say $0<\omega_1<\omega_2<\dots<\omega_M$:
$$
\frac{dz}{dt}= (\alpha-|z|^2)z+\imath\,\omega z+\ep\sum_{n=1}^M c_n\exp(\imath\,\omega_n\,t).
$$
Where $\ep$ is the combined forcing magnitude relative to other model parameters.
The complex coefficients in this trigonometric polynomial are
$$
c_n=\rho_n\exp(\imath\,\xi_n),
$$
and the variable $z$ may also be written in polar coordinates as
$$
z=r\exp(\imath\,\theta).
$$
With these notations, the equation for $z$ becomes
$$
\frac{dr}{dt}=(\alpha-r^2)r+\ep\sum_{n=1}^M \rho_n\cos(\omega_n\,t+\xi_n-\theta)
$$
and
$$
\frac{d\theta}{dt}=\omega+\ep\sum_{n=1}^M
\frac{\rho_n}{r}\sin(\omega_n\,t+\xi_n-\theta).
$$
This system has rich phase-locking properties. Let us choose one mode, say $\omega_N$ and set
$$
\theta=\omega_N\,t+\phi+\xi_N.
$$
Then
$$
\frac{dr}{dt}=(\alpha-r^2)r+\ep\sum_{n=1}^M
\frac{\rho_n}{r}\cos((\omega_n-\omega_N)\,t+(\xi_n-\xi_N)-\phi)
$$
and
$$
\frac{d\phi}{dt}=\omega-\omega_N+\ep\sum_{n=1}^M
\frac{\rho_n}{r}\sin((\omega_n-\omega_N)\,t+(\xi_n-\xi_N)-\phi).
$$
Averaging this system over $t$ gives for the second equation
$$
\frac{d\phi}{dt}=\omega-\omega_N+\ep \frac{\rho_N}{r}\sin(-\phi).
$$
Therefore, as long as the locking condition
$$
\bigg|\frac{\omega-\omega_N}{\rho_N\ep}r^*\bigg|<1
$$
is satisfied where $r^*$ is the equilibrium of the radial components, the oscillator will lock onto the $N$-th mode.

%We may plot this as output frequency
%($\lim_{t\to\infty}\theta(t)/t$) vs. $\omega$.
\subsection{Mutual Phase-locking of STNOs}
Consider now an array of $M$ STNOs with frequencies, say $0<\omega_1<\omega_2<\dots<\omega_M$:
$$
\frac{dz_i}{dt}= (\alpha-|z_i|^2)z_i+\imath\,\omega_i z_i+\ep\sum_{n=1}^M c_{i,n} z_n.
$$
The variables $z_i$ may be written in polar coordinates as
$$
z_i=r_i\exp(\imath\,\theta_i).
$$
With this notation, the equations for $z_i$ become
$$
\frac{dr_i}{dt}=(\alpha-r_i^2)r_i+\ep\sum_{n=1}^M c_{i,n}r_n\cos(\theta_i-\theta_n)
$$
and
$$
\frac{d\theta_i}{dt}=\omega_i+\ep\sum_{n=1}^M\frac{c_{i,n}r_n}{r_i}\sin(\theta_n-\theta_i).
$$
Let us look at one mode, say $\omega_N$ and assume certain symmetry, $c_{i,j}=c_{j,i}$. Mathematical analysis not carried out here shows that the amplitudes of both modes approach the same values, $r_i\rightarrow r^*$, unless one of the oscillators has zero initial value. In this case the difference of phases, $\Delta=\theta_i-\theta_N$ satisfies:
$$
\frac{d\Delta}{dt}=\omega_i-\omega_N-2\ep c_{i,N}\sin{\Delta}
$$
As before, we see that the system phase-locks if the condition
$$
\bigg|\frac{\omega_i-\omega_N}{2\ep c_{i,N}}\bigg|<1
$$
is satisfied.

\end{document}